\documentstyle[11pt,conf_iap,psfig]{article}
\newcommand{\kms}{km.s$^{-1}~$}
\begin{document}
\centerline {\bf EROS DIFFERENTIAL STUDIES OF CEPHEIDS IN THE MAGELLANIC CLOUDS : }

\centerline {\bf STELLAR PULSATION, STELLAR EVOLUTION \\ AND DISTANCE SCALE.} 
 
\centerline{ J.P. Beaulieu$^{1,2}$, D.D. Sasselov$^{1,3}$ }
\centerline{        beaulieu@astro.rug.nl, sasselov@cfa.harvard.edu }
\centerline{        1 Institut d'Astrophysique de Paris, CNRS, 98bis Boulevard Arago,F--75014 Paris, France. }
\centerline{        2 Kapteyn Laboratorium, Postbus 800, 9700 AV Groningen, The Netherlands.  }
\centerline{        3 Harvard-Smithsonian Center for Astrophysics, Cambridge, MA 02138, USA. }

\bigskip

\centerline {\bf Abstract } 
We present a differential study of 500 Magellanic Cepheids with 3 million measurements
obtained as a by-product of the EROS microlensing survey. The data-set is unbiased and
provides an excellent basis for a differential analysis between LMC and SMC.
We investigate the pulsational properties of the Cepheids as a
function of metallicity and the constraints 
they give to stellar pulsation, stellar evolution theory and opacities.

Moreover, our empirical analysis shows a small metallicity effect on the period-luminosity relation which 
has a significant implication to recent distance estimates based on HST observations of Cepheids
in galaxies as far as the Virgo cluster. When the color shift of Cepheids due to  metallicity is interpreted as 
reddening in the determination of the  true distance modulus of a target galaxy a systematic 
error is made. Accounting for that error makes the low-$H_0$ values (Sandage et al.) higher and the 
high-$H_0$ (Freedman et al.) value lower, bringing those discrepant estimates 
into agrement around $H_0 \approx 70$ kms$^{-1}$Mpc$^{-1}$.

\section{Introduction}

Cepheids are young, intermediate mass ( typically $2-10 M_\odot$), bright periodic variable stars. 
These stars have left the main sequence and are in a post core hydrogen-burning and core
helium-burning phase.
Because of this evolutionnary stage,  they lie in an area of the HR diagram, 
the so-called 'instability strip', where their envelopes are instable to global radial
pulsations via the kappa mechanism.
Their period of pulsation is tightly correlated with 
their luminosity, and this period-luminosity relation (PL) is used 
as a corner stone in deriving local and extragalactic distances for many decades.
Most of the recent determinations of the Hubble constant $H_0$ are based on HST observations
of Cepheids as far away as the Virgo cluster and an assumed universal PL relation
calibrated in the LMC. Then, secondary distance indicators are used to connect to the large scale
smooth Hubble flow.

One of the exciting by-products of the microlensing survey experiments like EROS\cite{CecIAP} and 
MACHO\cite{WelIAP,CookIAP} has been
the observation in a systematic way of hundreds of thousands stars in the Magellanic Clouds.
High quality light curves with excellent phase coverage have been obtained. 
We present a status report of the analysis of $\sim$500 Cepheids in the LMC and in
the SMC based on observation performed by EROS.

Magellanic Clouds are a perfect place to study variable stars. 
We can consider them at the same distance in a given galaxy, and the reddening is small. 
Therefore the observed differences are intrinsic. Moreover, these two galaxies have different 
metallicities. We  have the opportunity to check whether observations, stellar pulsation 
and stellar evolution theories give a self-consistent picture for two low metallicities and 
different chemical content.
With the two unbiased samples of Cepheids in the LMC and in the SMC, we have
empirically searched for the influence of metallicity on the PL relation.

\section{Cepheids as a test of stellar physics, stellar evolution and chemical composition}

\subsection{EROS observations}

Observations were obtained using  a 0.4 m f/10 reflecting telescope and a mosaic of 16  buttable CCDs
covering an area of 1 degree x 0.4 degree centered in the bar of the LMC. With the first year
of EROS CCD observations (1991-1992) about  2500 images spanning over $\sim$130 days, were taken in two broad bandpass 
filters $B_E$ and $R_E$  centered respectively on 490 and 670 nm. 
We have obtained 80 000 light curves, with as many  as 48 points in a given night. 
During the 1993-1995 season about 6000 images have been taken of a field centered in the SMC with a 
pair of very similar filters ($B_{E2},~R_{E2}$). An accurate transformation between the two systems 
have been determined\cite{H02}. For the 1991-1992 campaign, 9 CCDs were operative, whereas 
15 were working for the 1993-1995 campaign.
 
We searched systematically the EROS database for variable stars using the modified periodogram 
technique\cite{MPG,UAI155}, and the AoV method\cite{SCW,SMC96}.
We delineated the instability strip in the color magnitude diagram to select Cepheids candidates.
Then, we performed a visual inspection of the light curves to exclude eclipsing binaries. 
The remaining 97 stars in the LMC and 450 in the SMC form our sample of Cepheids.
One in the LMC (P0=3.4438 days, P1 = 2.443 days) and one in the SMC  (P0=2.0964 days, P1 = 1.5435 days) are beat Cepheids. 
For the SMC sample, improved signal processing techniques will be applied to find more candidates
to be compared with the ones discovered by MACHO\cite{Alc95,WelIAP}. We can notice already that the period ratio P10
of the double mode found in the SMC is higher than the ones found in the LMC for the same period. 
It is of special interest to see if the stars that pulsate both in the second and the first overtone 
modes in the SMC has the same period ratio P21 as the LMC ones. 

\subsection{Classical Cepheids and s-Cepheids.}

Since the last decade, Fourier decomposition techniques are proving to be a powerful tool
when used for systematic studies of variable stars. They give a quantitative basis for the 
comparison of non-linear hydrodynamic models  and amplitude equation formalism results with
observations\cite{BuchIAP}. A Fourier decomposition of the form 
$X=X_0+\sum_{i=1}^{N} X_i \cos (i \omega_i t + \Phi_i)$  is adopted and the customary quantities
$R_{k1} = X_k / X_1$ and $\Phi_{k1} = \Phi_k - k \Phi_1$ are calculated. The amplitude ratio
$R_{k1}$ reflects the asymmetry of the variation, and $\Phi_{k1}$ the full width at half maximum
of the curve. Both $R_{k1}$  and $\Phi_{k1}$ are strongly affected by resonances between 
pulsational modes.
For example, it is well known  that the so-called Hertsprung progression of the changing 
form of Classical Cepheid light curves is due to a 2:1 resonance between the
self-excited fundamental mode and the second overtone around $\sim 10$ days\cite{Sim,BuchIAP}.
In the $R_{21}-P$ plane, it corresponds to a minimum at $\sim 10$ days, and a steep rise 
followed by $ \pi / 2 $ drop at $\sim 10$ days in the $\Phi_{21}-P$ plane.  
Thanks to hydrodynamical models, it has been understood  that the center of the resonance
corresponds to the tip of the $\Phi_{21}-P$ curve just prior to the steep drop, and at the minimum of
the $R_{21}-P$.
This has been studied in our Galaxy\cite{Sim}, in the LMC\cite{Pet87}, and in the SMC\cite{And88}.
The resonance center ($P0=10$ days, period ratio $P20 = P2/P0 = 0.5$) can be used in 
conjunction with linear pulsation theory to infer  the mass and the luminosity of these stars.

For a few decades, Cepheids have been classified in two groups, the Classical Cepheids
with their high amplitude, asymmetric and possibly bumpy light curves that follow
the Hertzsprung progression, and the  s-Cepheids with low amplitude and symmetric light variation. 
The mapping of the Galactic short period Cepheids\cite{An86,Po94,An96} shows that s-Cepheids 
follow different sequences than the classical ones in the $R_{21}-P$ and $\Phi_{21}-P$ planes.
The difference has been interpretated as a difference of stellar evolution status 
(different crossing of the HR diagram), 
difference in chemical composition, or as consequence of different pulsational mode. 

\begin{figure}
\centerline{\psfig{figure=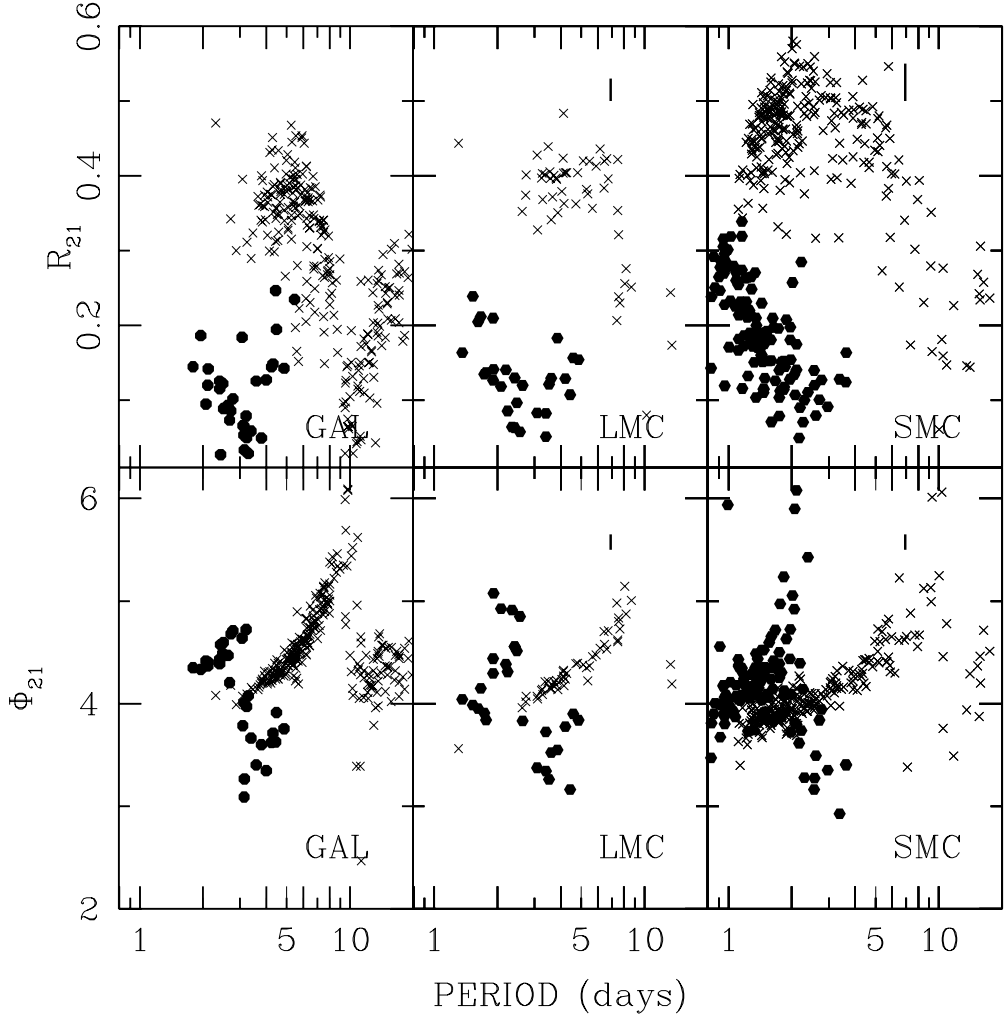,angle=0,height=14cm,width=14cm} }
\caption{ This figure gives the $R_{21}-P$, $\Phi_{21}-P$ plane for short period Galactic Cepheids
and EROS LMC and SMC Cepheids. S-Cepheids are plotted as filled diamonds, Classical Cepheids are plotted as crosses.  
Typical error bars are indicated in the upper part of each subfigure.}
\end{figure}

In figure 2 are plotted $R_{21}$ as a function of period ($R_{21}-P$), 
$\Phi_{21}$ as a function of period  ($\Phi_{21}-P$) for short period Galactic 
Cepheids\cite{Po94,An96}, and the EROS LMC\cite{LMC95} and SMC\cite{SMC96} Cepheids. 

A morphological separation based on the $R_{21}-P$ plane has been proposed for Galactic 
Cepheids\cite{An86}. We adopted it for LMC and SMC Cepheids, and classify as s-Cepheids the stars
lying in the lower part of the diagram. This separation in the $R_{21}-P$ plane 
is mirrored by a dichotomy in the PL diagram for LMC and SMC Cepheids (see figure 2 for the PL 
relation of SMC Cepheids).
Therefore we conclude that the difference is a consequence of different mode of pulsation, 
fundamental radial mode (F) for the Classical  Cepheid, and first overtone radial mode (1OT) 
for the s-Cepheids. 

\subsection{s-Cepheids}

In the 3 Galaxies, one can notice a rise in the $\Phi_{21}-P$ plane, followed by a sharp drop of $\pi / 2$
the so-called 'Z-shape'.
This feature is mirrored by a minimum in the $R_{21}-P$ plane.
The drop takes place at $P=3.2 \pm 0.2$ days in our galaxy\cite{Po94}, $P=2.7 \pm 0.2$ days in the 
LMC\cite{LMC95} and $2.2 \pm 0.2$ in the SMC\cite{SMC96}. It has been proposed that this drop is 
a signature of a 2:1 resonance between the first and the fourth overtone. Up to now, non-linear
survey of overtone pulsators have failed to reproduce this feature\cite{An93,An95,AlbIAP,Scha93}.
One should notice that it is surprising that this resonance could give birth to such a strong 
feature in Fourier planes because the fourth overtone is strongly damped. Alternative explanations, another class of 
fundamental mode pulsators with nearly-sinusoidal light curves for period greater than $\sim 3$ days, 
tangent bifurcation between first and second overtone pulsations, are not supported by the observed PL relation.

\subsection{Classical Cepheids}

The Hertzsprung progression and the 2:1 resonance between the second overtone and the fundamental 
mode for Classical Cepheids  is observed in the 3 galaxies. In our Galaxy, it is found to take place 
in the range 9.5-10.5 days\cite{MBM}. From MACHO observations\cite{WelIAP}, it is found to be between 
10.5-10.8 days in the LMC. With EROS data, we do not have enough Cepheids to constrain its position in 
the LMC. From SMC observations, it seems to take place in the range 10.5-11.5
days\cite{SMC96,And88}.

Figure 2 show that it is possible to define a common envelope for the Cepheids 
of the 3 galaxies before the 10 days resonance in the $R_{21}-P$ plane. Can the spread of values 
in $R_{21}$ show different amount of damping ? Is it  correlated with position in the instability strip ? 
It is also remarkable to notice that the Cepheids of the 3 Galaxies follow the same sequence in the $\Phi_{21}-P$
plane before the resonance. 
However, Cepheids of shorter period are  observed in the LMC than in the galaxy, and even  shorter ones are observed in the SMC.
Both Cepheids from the 3 Galaxies present similar acoustic properties : the envelopes that can be plotted in the Fourier planes
give the limits of pulsational properties of the stars. 

The metallicity affects the evolutionary tracks. For high metallicity, the blue loops for low mass stars do not cross 
the instability strip, whereas they do for low metallicities. 
With decreasing the metallicity, the lower limit for the mass of a star to become a Cepheid decrease.
It gives a strong constraint about the extension of the blue loops in the HR diagrams for different metallicity. 

\begin{figure}
\centerline{\psfig{figure=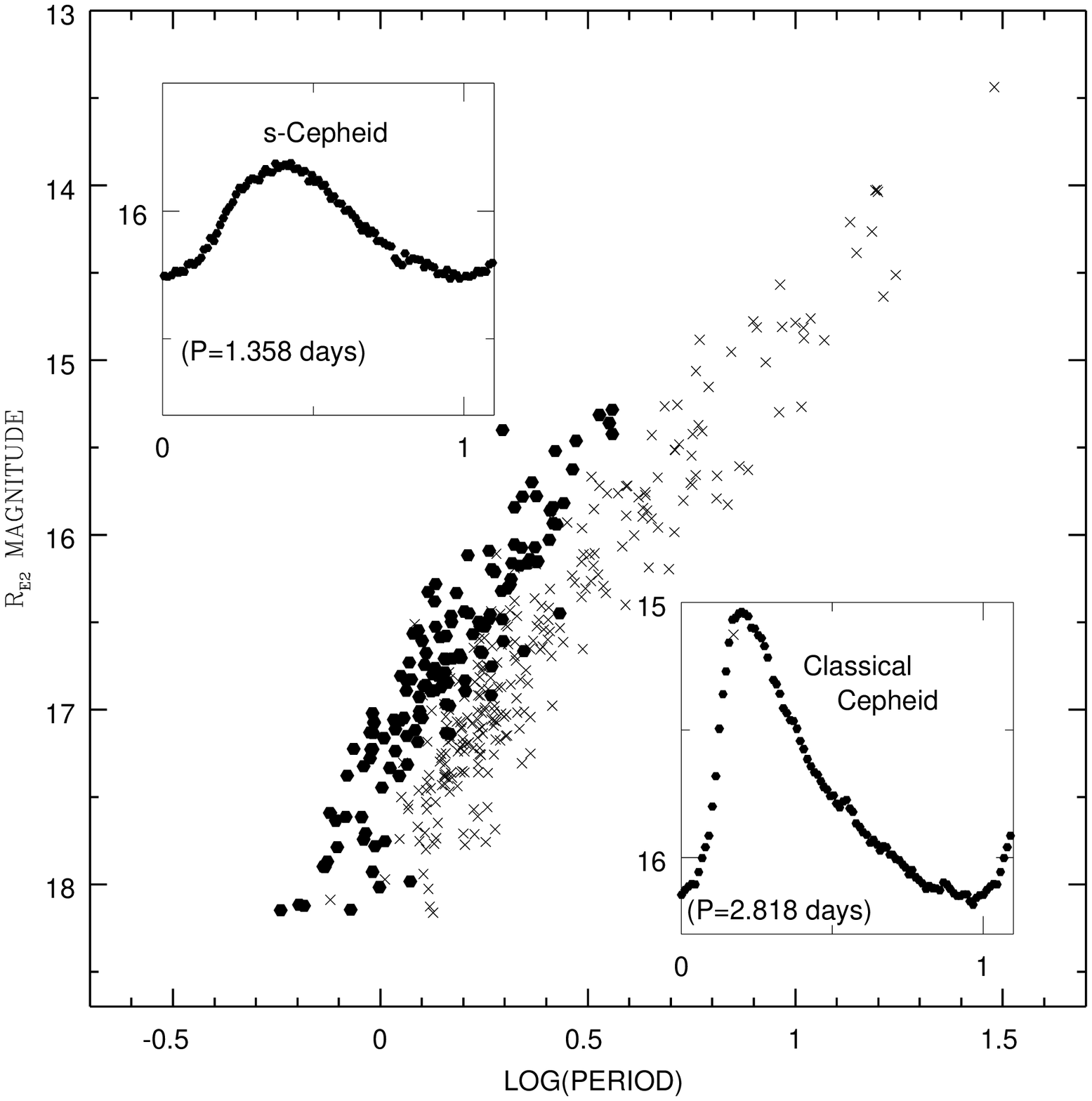,angle=0,height=8cm,width=8cm}
            \psfig{figure=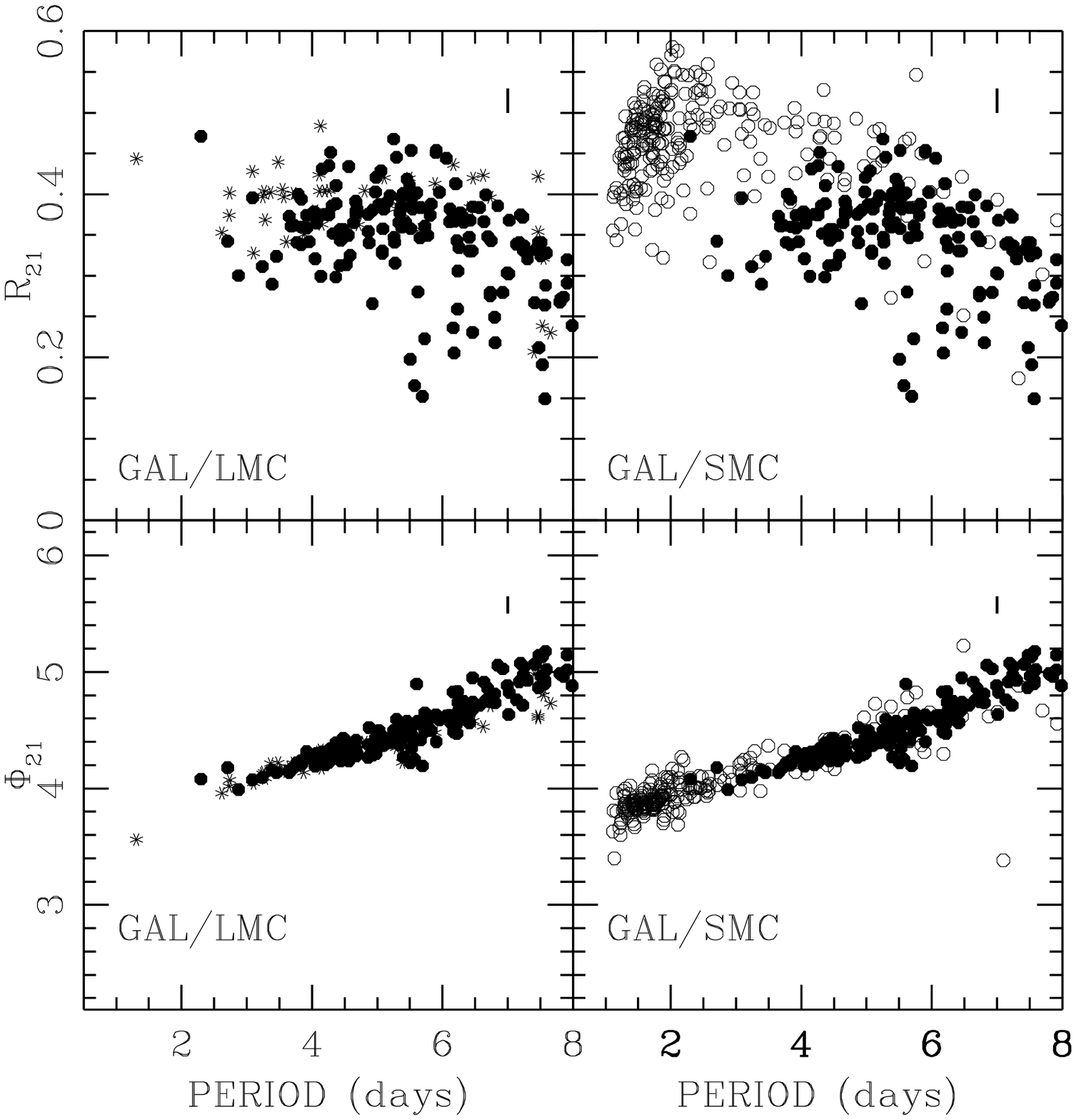,angle=0,height=8cm,width=8cm} }
\caption{The Period-Luminosity relation for SMC Cepheids from the EROS database. Classical
Cepheids (fundamental mode pulsators) are plotted as crosses, and s-Cepheids (overtone pulsators)
are plotted as dots. Two examples of classical (lower right) and s-Cepheids (upper left) 
light curves are shown in the figure. The same scale for the amplitude has been adopted. 
The two types of Cepheids can be isolated on morphological basis, and follow different 
PL relation.  In the right figures  are plotted the $R_{21}-P$, $\Phi_{21}-P$
for fundamental pulsators in the Galaxy (dots), the LMC (asterix), the SMC (circle). 
Typical error bars are indicated in the upper part of each subfigure.}
\end{figure}

\subsection{Discussion about resonance constraints}

When going to lower metallicity the $\sim10$ days resonance occurs at longer period.
For overtone pulsators, the $Z-shape$ takes place at shorter period when decreasing metallicity.

From a theoretical point of view, one has to keep in mind that when going to lower metallicities, 
the evolutionary models will increase the luminosity at fixed mass. This increase of the mass 
will lead to a diminution of the calculated period ratios. However, when going to lower Z, the opacity bump 
that drives the pulsation will be smaller, and therefore will increase the period ratios for a 
fixed mass and luminosity. The final results, the observed position of a resonance center, will 
be a combination of these two antagonists effects (plus a non-linear shift, which is of the level of a 
few tenth of a percent in period ratio for the 10 days resonance, for a solar composition and metallicity\cite{MBM}).

\subsection{Using resonance constraints for linear stability envelopes calculations}

First of all, we want to stress that up to now, there is no correct non-linear model
for Beat Cepheids. Therefore, one cannot know the influence of non-linear shift compared to the results
obtained by linear theory. One earlier study\cite{Puz96} has shown that with the current state of art, 
if one allows a 1 \% shift on period ratio due to a non-linear coupling effects, linear stability analysis of 
hydrostatic envelopes of Beat Cepheids gives weak constraints about mass-luminosity relation. If one
plots mass as a function of period ratio for a set of models to find the mass and the luminosity
of a given beat Cepheid and  assume that the non-linear period shift is handled, the nearly horizontal
nature of the curves shows that it is difficult to get reliable mass estimates for beat Cepheids.
The situation is worse for long period beat Cepheids than for short period ones.
However, some efforts have been concentrated on them\cite{MBM,Sim94,Chris95,WelIAP,Morgan96}.

The presence of a resonance in the envelope of Cepheids gives an anchor for linear pulsation
theory to infer the mass and the radius of these stars.  Since the introduction of the new OPAL 
opacities in 1991, masses  determined by Baade-Wesselink analysis, stellar evolution theory and 
linear pulsation theory are in agreement for the solar metallicity\cite{MBM,Kan94}.
The new data obtained by EROS and MACHO in the LMC and in the SMC poses a challenge for stellar 
pulsation and stellar evolution, by inviting to investigate lower metallicities.

One previous survey\cite{Puz96} showed that the implication of the 2:1 resonance for fundamental
Cepheids around 10 days and the alleged 2:1 resonance around 3 days for overtone pulsators are difficult 
to reconcile with the envelope models of low-metallicity (Z=0.01 and Z=0.004) Cepheids. When going to
lower metallicity, the masses derived by our linear calculations are too small.
To summarize, a systematic survey of linear stability analysis of Cepheid envelopes have been done\cite{Puz96}.
The hydrostatic models are computed with a 200 zones mesh using the OPAL opacity table 
(OPAL93\cite{OPAL96})  and Alexander Ferguson opacities\cite{AF94} for the chemical composition
X=0.70 etc Z=0.02, Z=0.01, Z=0.004.
We find by iteration the mass (M) and the luminosity (L) that are compatible with a given resonance 
constraint. We computed the models at different distances from the linear blue edge (0, 200 K and 400K)
for the corresponding M and L. 
To analyses the deficiency of our previous calculations\cite{Puz96}, we performed a new survey with the
latest OPAL opacities (OPAL95) to test the sensitivity to various effects.\\

\noindent (0) The use of the new OPAL opacities (OPAL95 instead of OPAL93) didn't solve the problems. There was a small 
increase of masses of a few percent thanks to the inclusion of new metals, and therefore an increase of the driving
'metal bump'. For Z=0.02, we have good agreement between the masses given by the stellar pulsation
theory and the masses given by the evolutionary calculations of the Geneva group\cite{SSMM} and masses determined by Baade 
Wesselink analysis. To summarize, $Z=0.03$ with OPAL93 is more or less equivalent to $Z=0.02$ with OPAL95.
However, the discrepancy for LMC and SMC Cepheids mass determinations has not been removed. 
A gain of $\sim$50 \% in mass for SMC Cepheids, and $\sim$30 \% for LMC Cepheids would be needed to be in agreement with the 
evolutionary calculations.

\begin{figure}
\centerline{\psfig{figure=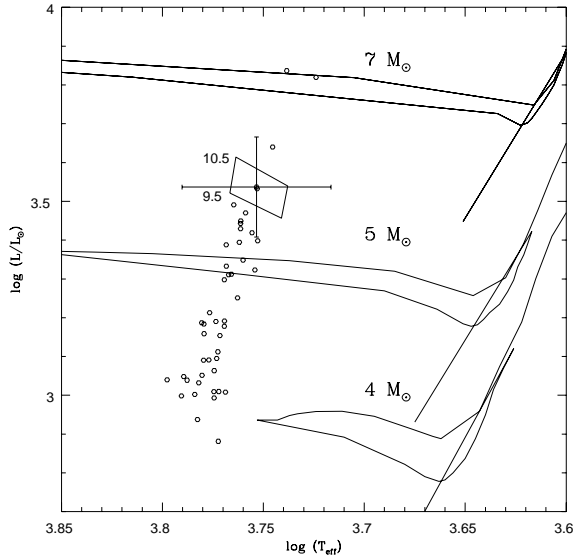,angle=0,height=8cm,width=8cm}  }
\caption{We give the $\log(L/L_\odot)-\log(T_{eff})$ diagram for LMC fundamental pulsators (plotted as dots). 
We derived individual reddening, then transform to standard system and apply a temperature
scale\cite{Chio93}. We assume a distance modulus of $18.5 \pm 0.13$ mag to the LMC\cite{Pan91}. 
The error bars  ($1 \sigma $) are plotted on a 10.2 day Cepheids.  
They are dominated by the color equation transformation\cite{BIN95}.
One should note that when deriving individual reddening we over-correct for the position in
the instability strip. Therefore it appears slightly narrower than it is. 
The evolutionary tracks are from the Geneva group with Z=0.008. The box comes from 
linear stability analysis of hydrostatic models of resonant Cepheids with period 9.5 and 10.5 days
(with Z=0.01 and OPAL96 opacities) at 0, 200 K from the linear blue edge. The masses for a 10.5 days Cepheid
at the 200 K from the linear blue edge is about $4.9 M_\odot$ to be compared with the evolutionnary track of $4 M_\odot$ and $5 M_\odot$.}
\end{figure}

\noindent (1) We increase the number of zones in our mesh up to 800. The variations we found are smaller than a few tenths of a
percent in period ratio.

\noindent (2) We included convection following a classical mixing length theory. Surprisingly, the situation worsen, and we obtained masses
even smaller.

\noindent (3) We made a comparison between the two available mixture available with OPAL tables, Grevesse91, and Grevesse93.
The main difference is a 5 \% increase of the opacity in the range $log(T) = 5.3-6$K corresponding to the inclusion of more 
metals, and changes of abundances between the two mixtures, that gets an increase of the mass of about 5 \% for a given model.
No significant differences have been found between OPAL and OP opacities when using linear pulsation calculation\cite{Sim94}. 
It would be necessary to get reliable information about the chemical composition of Cepheids in the Magellanic Cloud and in 
the Galaxy, in order to adjust opacity tables and analyze its influence on the pulsation.
Moreover, we have to underline that we use a metallicity of $Z=0.01$ for  LMC (which is a rather high estimate), 
whereas we use Z=0.004 for SMC (which is a rather low estimate). 
 
\noindent (4) We performed an extensive survey with enhanced opacities in different temperature ranges to try to increase the masses 
of our models. We found that the most sensitive region  is (as it could have been expected from inspection of weight functions
and our knowledge of stellar pulsation) in the range $\log T = 5-5.3$ (metal bump region) 
where a crude increase of 20 \% gives a 30 \% increase of the mass for Z=0.004. One should note that it is more or 
less equivalent to increase the metallicity of the models. So it is an 'ad-hoc' or 'cooking' experiment.

\noindent (5) In the first survey\cite{Puz96}, it has been assumed that the resonance takes place for the 3 galaxies 
between 9.5-10.5 days, and resonant models for these periods have been computed and discussed. It would be more appropriate
to compare a 10 days Cepheid in our Galaxy, with a 10.5 days in the LMC and a 11 days in the SMC. In table 1 are given some 
models computed by the iterative code. One should keep in mind that the actual fundamental blue edge is shifted to lower 
temperature by about 100 K, because the first overtone mode is linearly unstable. Far from the blue edge, the results of the 
calculations are less reliable because of the possible influence of convection that is ignored by the code. 
At 200 K from the linear blue edge, one can notice that a mass of 5.9 $M_\odot$ is obtained for the Galaxy whereas, 
$4.9 M_\odot$ is obtained for the SMC, and $4.3 M_\odot$ for the SMC.

\begin{table}
\caption[]{ Models computed with OPAL96 and the iterative process described in the text\cite{Puz96,MOD96}.
For each galaxy, we computed a resonant Cepheid of a given period at the blue edge, at the blue edge 
+200 K, and +400 K. Masses, Luminosities and effective temperatures are given for all the models.}
\begin{flushleft}
\begin{tabular}{llllllllllll}
\hline
              & mass & Lum  &  Teff& & mass & Lum  &  Teff&  &mass & Lum  & Teff \\
              &  ($M_\odot$) & ($L_\odot$) & (K) &  & ($M_\odot$) & ($L_\odot$) & (K) 
              & &  ($M_\odot$) & ($L_\odot$) & (K) \\
\hline
GAL 10 days   & 6.06 & 4243 & 5648 & $\|$ & 5.90 & 3721 & 5489 & $\|$ & 5.71 & 3241 & 5331 \\
LMC 10.5 days & 4.57 & 4109 & 5802 & $\|$ & 4.88 & 3849 & 5634 & $\|$ & 4.96 & 3464 & 5469 \\
SMC 11 days   & 3.63 & 3882 & 5873 & $\|$ & 4.26 & 3927 & 5704 & $\|$ & 4.51 & 3635 & 5537 \\
\hline
\end{tabular}
\end{flushleft}
\end{table}

There is still a problem in reconciling the Cepheid mass determined by our models for the LMC and SMC metallicity using 
the resonnance contraint given by the observations of the $\sim 10$ days resonnance despite our new efforts. A slight
increase of the luminosity of the evolutionnary calculations, or a small increase of the masses derived by linear stability 
analysis of hydrostatic models is needed, or an increase of the opacity tables in the 'metal bump region', or the three...
We want to underline again that envelope stability analysis are very sensitive to opacities, whereas evolutionary calculations 
are less. The inclusion of new metals in the opacity tables would increase a bit the metal opacity bump, and therefore may 
improve a situation, that is already not so bad !

\section{Cepheids and the distance scale}

\subsection{Cepheid distance determination}

The long-term $HST$ distance scale programs\cite{F94,Tan95,San94} have all adopted the same 
method\cite{MF91,FM96} based on two-color photometry to determine reddening corrected 
Cepheid distances.
It is assumed that the Cepheid PL relation is universal, and that the Cepheids from the calibrating
set and from the target galaxy have the same intrinsic colors.
The slopes of a calibrating set of LMC Cepheids are calculated. The PL relation 
of the Cepheids from the target galaxy is slided against it to derive apparent distance modulus
in each band. A true distance modulus of $\mu_{LMC}=18.5$mag, a mean reddening of 
$E(B-V)=0.10$ and a Galactic extinction law with $R_V=3.3$ are adopted for the LMC.
Then, using the multicolor apparent distance moduli and a Galactic extinction law, the total 
mean reddening and the true distance modulus of the target galaxy are determined.

However, theory\cite{Stothers88,St95} predicts a small metallicity effect on the PL relation 
as a results
of stellar pulsation through the dependence of period on mass and radius; of stellar 
evolution through the mass luminosity relation; and of stellar atmospheres theory through the relation 
between effective temperature, absolute magnitude in bandpasses and bolometric correction.
As a summary theory predicts that the slopes of the PL relation are independent of Z, only the zero-point
is affected, and the metallicity effect depend on band pass. It is predicted to 
be small, $\delta M_V=-1.1 \delta Z$. However, {\bf if one interpretes the color shift due to metallicity 
as reddening in deriving the true distance modulus of a target galaxy with the method describe above, 
then one makes a systematic error of $\bf \delta \mu  = \delta M_V + R_V \delta(B-V)_0$, 
$\bf \delta \mu = 28.6  \delta Z $, according to theory\cite{Stothers88}} ($\delta(B-V)_0$ is the color change 
due to metallicity).

Observational studies were made since the 70s. For example, an intrinsic color shift between LMC and SMC Cepheids has been 
clearly pointed out\cite{M79}. An empirical search for a metallicity effect\cite{FM90,Gould94} in three fields of M31 with 
36 Cepheids and 152 BVRI measurements have led to ambiguous results : Freedman \& Madore\cite{FM90}
claimed that there is no significant effect. Gould\cite{Gould94} reanalyzed their data with a better statistical 
treatment taking into account the high degree of correlation between the measurements and found an effect.
However, due to the number of observations, he did not solve for a wavelength dependent metallicity 
effect.
The current status is that the influence of metallicity is  considered to be insignificant\cite{F94,San94}, 
or negligeable\cite{Tan95}, and is ignored in the derivation of extragalactic distances with
$HST$.

We used the EROS Cepheids data from LMC and SMC to perform an empirical test for metallicity effect\cite{H01,H02}.
We have high quality, excellent phase coverage light curves for classical Cepheids and s-Cepheids. Since 
they pulsate in different modes, they follow different PL relations. In the LMC we keep 51 fundamental pulsators 
and 27 first overtone pulsators, and 264 fundamental pulsators and 141 first overtone pulsators in the SMC.
So, we have two unbiased samples of Cepheids that fill densely the period-luminosity-color (PLC) space, with known difference
in metallicity $\delta [Fe/H]_{LMC-SMC} = 0.35$\cite{Luck92}.

\subsection{The method}

Our method\cite{H01,H02} will be applied independently to classical Cepheids and s-Cepheids.
First we compute wavelength dependent slopes for LMC and SMC Cepheids. They are the same 
in both galaxies within the uncertainties. 

We  search for a metallicity effect that depends upon band pass and
model the data in the PLC plane taking into account the high degree of correlation between
the measurements\cite{Gould94}. The assumption of our model are constant PL slope with metallicity,
and no depth dispersion with the LMC sample. We adopted an LMC 
mean reddening of $E(B-V)=0.10$ and a Galactic extinction law with $R_V=3.3$.
The model then has 12 parameters, which are : 
two linear fits in the PLC space (slopes $\beta_i$ and $b_i$, and the zero points $\alpha_i$
and $a_i$), the distance difference $\gamma_1$,  the relative reddening difference $\gamma_2$
and a metallicity dependence $\gamma_3^i$ (i=1,2 are the two colours $B_E$ and the $R_E$).
We note that the observed mean magnitude in the $i$ band for a Cepheid $p$, 
with period $P_p$, is $Q_{i,p}$. The residuals for the SMC sample are then :  

\begin{equation}
X_{i,p}^3=Q_{i,p}-(\alpha_i+\beta_ilogP_p+\gamma_1+\gamma_2R_i+\gamma_3^i),
\end{equation}
$$
X_{i,p}^4=Q_{i,p}-[a_i+b_i(Q_{1,p}-Q_{2,p}+\gamma_2(R_2-R_1)~~~~~~~~~~~~~~~~
$$
\begin{equation}
~~~~~~~~~~~~~~+\gamma_3^1-\gamma_3^2)+\gamma_1+\gamma_2R_i+\gamma_3^i],
\end{equation}

 with $p=1,...,N(n)$ being the number of Cepheids. 
The form of the covariance matrices of the data, the $\chi^2$ minimization and the iteration procedure
are described in \cite{H02}. Unlike Gould (1994), we solve for a wavelength dependent metallicity effect
and constrain the individual and sample reddenings of the Cepheids by the independently known foreground 
reddenings of LMC (0.06) and SMC (0.05). Details of the method are given elsewhere\cite{H01,H02}.

We applied the technique to the Classical and the s-Cepheids  independently and obtained the same
results. We derived a correction term due to the  metallicity dependence to the distance 
determination method presented in 3.1.

$$\delta \mu = (25.4^{+6}_{-12})[Z-0.009]$$

This metallicity term is valid in the spectral region covered by EROS filters 
and can be applied to V, I photometry as shown by the color equation transformation
$(V-I) = 1.02 (B_E-R_E),~\sigma=0.02$mag.

\subsection { Effect on $H_0$ determination }

Numerous efforts to determine the Hubble constant $H_0$ have led to dichotomous results,
depending on the choice of the cosmological ladder used\cite{Fuk93}. However, both
the longer distance scale ($H_0 \approx 50$ kms$^{-1}$Mpc$^{-1}$), 
and the shorter distance scale ($H_0 \approx 80$ kms$^{-1}$Mpc$^{-1}$) rely strongly on 
the Cepheid PL relation calibrated in the LMC and the method described in section 3.1.
 
Using HST $VI$ photometry of Cepheids in M100\cite{F94}, $H_0=80\pm17$ \kms $Mpc^{-1}$ has been derived.
M100 Cepheids are metal rich compared to LMC ones. Adopting [Fe/H]=+0.1\cite{Zar94}, within the 
uncertainty range of our metallicity effect, this leads to $H_0=76-70$ \kms $Mpc^{-1}$. 
Recently, Cepheid distances to a few more galaxies less metal rich than M100, and even with 
comparable metallicity to LMC (outer region of M101) have been reported by the same group\cite{FM96}. It brings 
down their initial value obtained for M100, and nicely confirms the effect of metallicity on the Cepheid
distance scale, since their new distances are less affected by it. 

HST $VI$ photometry of Cepheids have been obtained in M96\cite{Tan95} and $H_0=69\pm8$ \kms $Mpc^{-1}$.
This galaxy has an abundance estimate of $[Fe/H]=-0.02$. 
This imply a small correction in the Cepheid distance that leads to $H_0 \approx 66 $ \kms $Mpc^{-1}$.

Using HST $VI$ photometry of Cepheids in IC4182 and NGC5253\cite{San94}, $H_0=55\pm8$ \kms $Mpc^{-1}$
has been derived. These two galaxies are metal-poor compared to the LMC. 
We adopt [Fe/H]=-1.3\cite{Pagel92,Saha96}, caution on the large uncertainties of this metallicity
determination, and the correction leads to  $H_0=59-68$ \kms $Mpc^{-1}$.

 The metallicity dependence we found from the LMC/SMC analysis brings all
the derivations of $H_0$ to good agreement around $H_0 \approx 70$ \kms $Mpc^{-1}$. 

\vskip 2.0 truecm
{\sl It is a pleasure to thank Robert Buchler, Marie-Jo Goupil and Zoltan Koll\`ath 
for very pleasant and useful discussions. The modelling part has been done in collaboration with them. 
We thank Roger Ferlet and Alfred Vidal-Madjar and the members of the EROS team for their continuous support.
This work is based on observations held at ESO La Silla.}

\vfill
\end{document}